\documentclass[a4paper,fleqn]{cas-sc}

\usepackage[numbers,sort&compress]{natbib}

\def\tsc#1{\csdef{#1}{\textsc{\lowercase{#1}}\xspace}}
\tsc{WGM}
\tsc{QE}

\begin{document}
\let\WriteBookmarks\relax
\def\floatpagepagefraction{1}
\def\textpagefraction{.001}

\shorttitle{Action potentials and solitons}    

\shortauthors{J.Engelbrecht, K.Tamm, T.Peets}  

\title [mode = title]{Action potentials and solitons}  



%

\author[1,2]{J\"uri Engelbrecht}

\cormark[1]


\ead{je@ioc.ee}


\credit{Conceptualization of this study, initial manuscript draft, conclusions}

\affiliation[1]{organization={Tallinn University of Technology},
            addressline={Ehitajate tee 5}, 
            city={Tallinn},
            postcode={19086}, 
            country={Estonia}}

\author[1]{Kert Tamm}[orcid=0000-0002-2455-8258,]


\ead{kert.tamm@taltech.ee}


\credit{Initial manuscript draft, numerical simulations, preparing figures, analysis of data, conclusions}

\affiliation[2]{organization={Estonian Academy of Sciences},
            addressline={Kohtu 6}, 
            city={Tallinn},
            postcode={10130}, 
            country={Estonia}}

\cortext[1]{Corresponding author}


\author[1]{Tanel Peets}[orcid=0000-0002-3632-2664,]

\ead{tanel.peets@taltech.ee}

\credit{Analysis of data, conclusions, technical editing of manuscript}


\begin{abstract}
During the last decade, the notion of solitons has been mentioned in neuroscience related to  the propagation of action potentials (AP). In this paper, based on many studies in mathematical physics and neuroscience, the clear differences between the APs and solitons are summarised. It is stressed that the physics of the emergence of APs and solitons is fundamentally different. The numerical examples collected from earlier studies demonstrate the differences in the process of generation and interaction of corresponding waves explicitly. It is also noted that although the longitudinal mechanical waves in biomembranes can be described by the Boussinesq-type equation, the time-scale of emerging solitons from an arbitrary initial excitation exceeds the time-scale of the usual AP existence considerably. 
\end{abstract}


\begin{highlights}
\item The emergence of action potentials and solitons is governed by physically distinct mechanisms.
\item The interaction mechanisms of action potentials and solitons are different.
\item The terminology of mathematical physics must be used properly.
\end{highlights}

\begin{keywords}
action potential \sep soliton \sep terminology 
\end{keywords}

\maketitle

\section{Introduction}\label{Intro}

Much research has been done, especially over the last century, to understand the processes in neurons.  An important step in neuroscience was made by A.~Hodgkin and A.~Huxley, who explained the formation of the action potential (AP) by the ion mechanism. The Hodgkin-Huxley (HH) model describes an asymmetric solitary pulse with an overshoot as the main signal in nerves \cite{Hodgkin1952}, and the experiments have proved the shape of such a wave. In general terms, the HH model describes the waves in an active medium, where the ion currents through the biomembrane support the generation of the AP. 

Another important avenue of research is related to solitons: solitary pulses in a nonlinear dispersive medium, discovered in fluid dynamics and later in many other media (solids, plasma, gases). An emotional description of solitons in a narrow water channel in Edinburgh by John Scott Russell in 1834 marks the beginning of studies. The next major step is related to evolution of computers and Fermi, Pasta, Ulam and Tsingou's studies in 1955 in Los Alamos related to the equipartition of energy in nonlinear lattices. The breakthrough was made about a decade later by N.~Zabusky and M.~Kruskal in 1965 \cite{Zabusky1965}, who studied the process governed by the Korteweg-deVries (KdV) equation -- a model one-wave equation including nonlinear and dispersive effects. The KdV equation has solitary-type solutions with special properties that are characteristic of elementary particles. That is why Zabusky and Kruskal invented the term ``soliton'', where the suffix ``on'' emphasises the similarity to electrons, baryons, etc. Later, it was discovered that many equations have soliton-type solutions (Sine-Gordon, nonlinear Schrödinger, Boussinesq, etc) and there are many practical applications in fiber optics, material physics, fluid dynamics, etc.

The stability of soliton-type waves may raise a question whether an AP can be described as a soliton. Scott \cite{Scott1999} gave an explicit explanation that the processes of generation of APs in neurons and solitons in nonlinear and dispersive media are physically different and should not be mixed up. 

The experimental studies have revealed that, besides the AP, there are other physical effects accompanying the propagation of the AP. These effects are (i) mechanical, involving the deformation  of the biomembrane and axoplasm and (ii) thermal, i.e., change of the temperature. These studies were certainly known to Hodgkin, who has stated \cite{Hodgkin1964a}: ``In thinking about the physical basis of action potential, perhaps the most important thing to do at the present moment is to consider whether there are any unexplained observations which have been neglected in an attempt to make experiments fit into a tidy pattern.'' This statement was followed in many studies. An interesting proposal was made by Heimburg and Jackson  \cite{Heimburg2005} describing the deformation (density change) of the biomembrane. They proposed a Boussinesq-type equation for describing the longitudinal density change of the biomembrane. The Heimburg-Jackson (HJ) model involves a displacement-type nonlinearity and a dispersive term due to the elasticity of the microstructure (the lipid bilayer) of the biomembrane. This equation has a soliton-type solution. According to Heimburg and Jackson, such a soliton plays the most important role in the neural signal. This idea has found supporters, and now the soliton concept is  sometimes mentioned in neuroscience. However, it seems that  often the notion of a soliton in neuroscience is used without the proper analysis, although Heimburg and Jackson \cite{Heimburg2005} stressed explicitly that their study is on the propagation of solitons in cylindrical membranes. The present essay is written with the aim of explaining the difference between a solitary wave and a soliton. 

Some examples (without references) illustrate the need for such an explanation. In one of the reviews of a submitted paper of the present authors, it was said:  ``Especially since the authors observe that pulses propagate with stable profiles, have well-defined velocities, annihilate upon head-on collision, exhibit threshold behaviour, and show refractory periods, we consider these features to be hallmarks of soliton-like excitations in nonlinear systems.''  

A preprint about the possible mathematical models of the AP states: ``There is no doubt that action potential is a soliton''. However, the author does not present a mathematical model of this ``soliton''. Here and there, the phrase ``nerve soliton pulses'' is used, and solitons are associated with the neural signal. We would like to stress the need for using terms of mathematical physics properly. 

In what follows, the differences between the concepts of solitary waves and solitons are described, illustrated by computational examples. Section \ref{Definitions} is devoted to definitions, and further on, Section \ref{Propagation} demonstrates the propagation effects. The mathematical model for the AP and the accompanying effects is based on the studies \cite{Hodgkin1952,Lieberstein1967,Raamat2021,Tamm2025,Tamm2026}. The generation and the propagation of solitons are based on the analysis of the Boussinesq-type equations \cite{Raamat2021,Peets2015,Christov2007,EngSalupTamm2011}. The conclusions are summarised in Section \ref{Final}.

\section{Definitions}\label{Definitions}
The action potential in axons is measured in many labs around the world, and its properties are well known, starting from Hodgkin and Huxley studies \cite{Hodgkin1952}, and described in many monographs (see, for example, \cite{Nelson2004,Raamat2021}, etc.). The AP is an asymmetric pulse (a solitary wave) characterised by specific features \cite{Nelson2004}:

\begin{itemize}
\item it is an all-or-nothing response: if the stimulus is below threshold, then there will be no response far from the stimulating point; if the stimulus is above threshold, then a travelling wave is created whose peak amplitude does not depend on the strength of the stimulus;
\item the AP propagates along the axon at a constant velocity dictated by the properties of the axon;
\item the peak of the AP is found independent of distance and preserves its shape; however, later experiments have shown a possible decrease in the amplitude, the amplification or the broadening of the AP; 
\item after the passage of an AP, the membrane potential has an overshoot and then slowly recovers;
\item before the resting potential is restored, no new AP can be generated, which means the existence of a refractory period;
\item two counterpropagating APs annihilate each other.
\end{itemize}

\begin{figure}[p]
  \centering
    \includegraphics[width=0.49\textwidth]{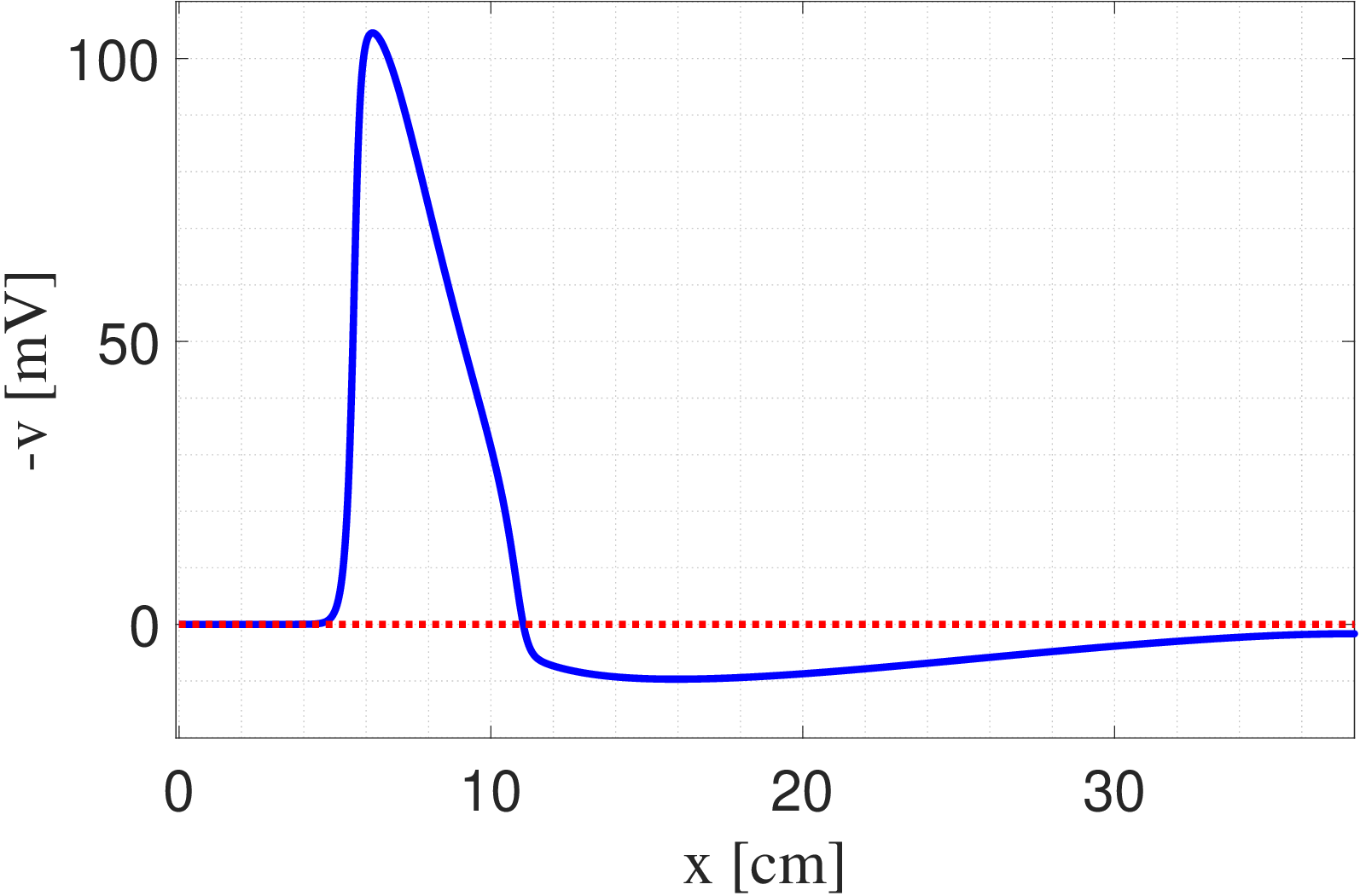}
    \includegraphics[width=0.49\textwidth]{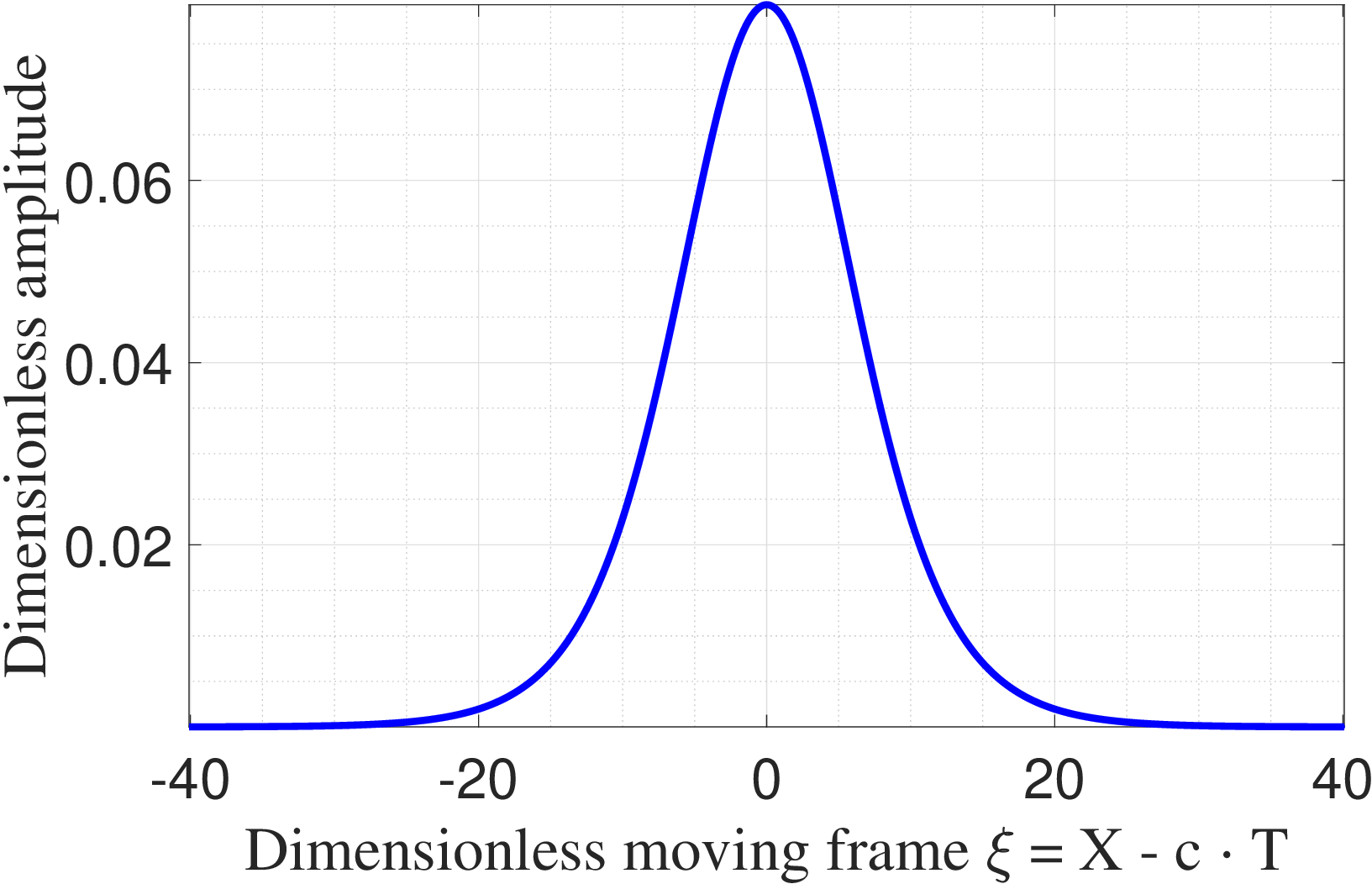}
    \caption{Left panel -- typical action potential (AP) (model eqs.~\eqref{LIB11} and \eqref{LIB21} \cite{Tamm2025}). Right panel --  soliton profile (see eq.~\eqref{HJsoliton} for analytical solution of improved Heimburg-Jackson model \eqref{iHJ})  (see  \cite{Engelbrecht2017} for details). Here (in eq.~\eqref{HJsoliton}) $X$ is dimensionless space, $T$ is dimensionless time, $c$ is normalised (dimensionless) velocity, and $\xi$ represents a moving frame of reference propagating with the soliton-type solution.}\label{fig1}
\end{figure}

A soliton is a solitary wave which can be generated in nonlinear dispersive media. Usually, its  classical definition is based on the KdV equation \cite{Zabusky1965,Ablowitz2011}. A soliton is:
\begin{itemize}
\item a nonlinear wave which maintains its shape,
\item propagates with a constant velocity depending on its amplitude,
\item restores its shape after a collision with another soliton, except for a phase shift.
\end{itemize} 
Depending on the input, solitons may have different amplitudes.
The typical profiles of these two types of solitary waves are shown in Figure \ref{fig1}.

It is important to note the different mechanisms of generation of an AP and a soliton. An AP, according to the HH paradigm, exists because the ion mechanism changes the ion concentration in axoplasm, i.e., an AP propagates in an active medium. A soliton is a wave in a conservative medium (i.e., no dissipation or energy inflow), and its generation is governed by the balance of nonlinear and dispersive effects. Despite clear differences, sometimes these two concepts are mixed up, as mentioned in Section \ref{Intro}.  

\section{Propagation effects}\label{Propagation}

\begin{figure}[p]
  \centering
    \includegraphics[width=0.49\textwidth]{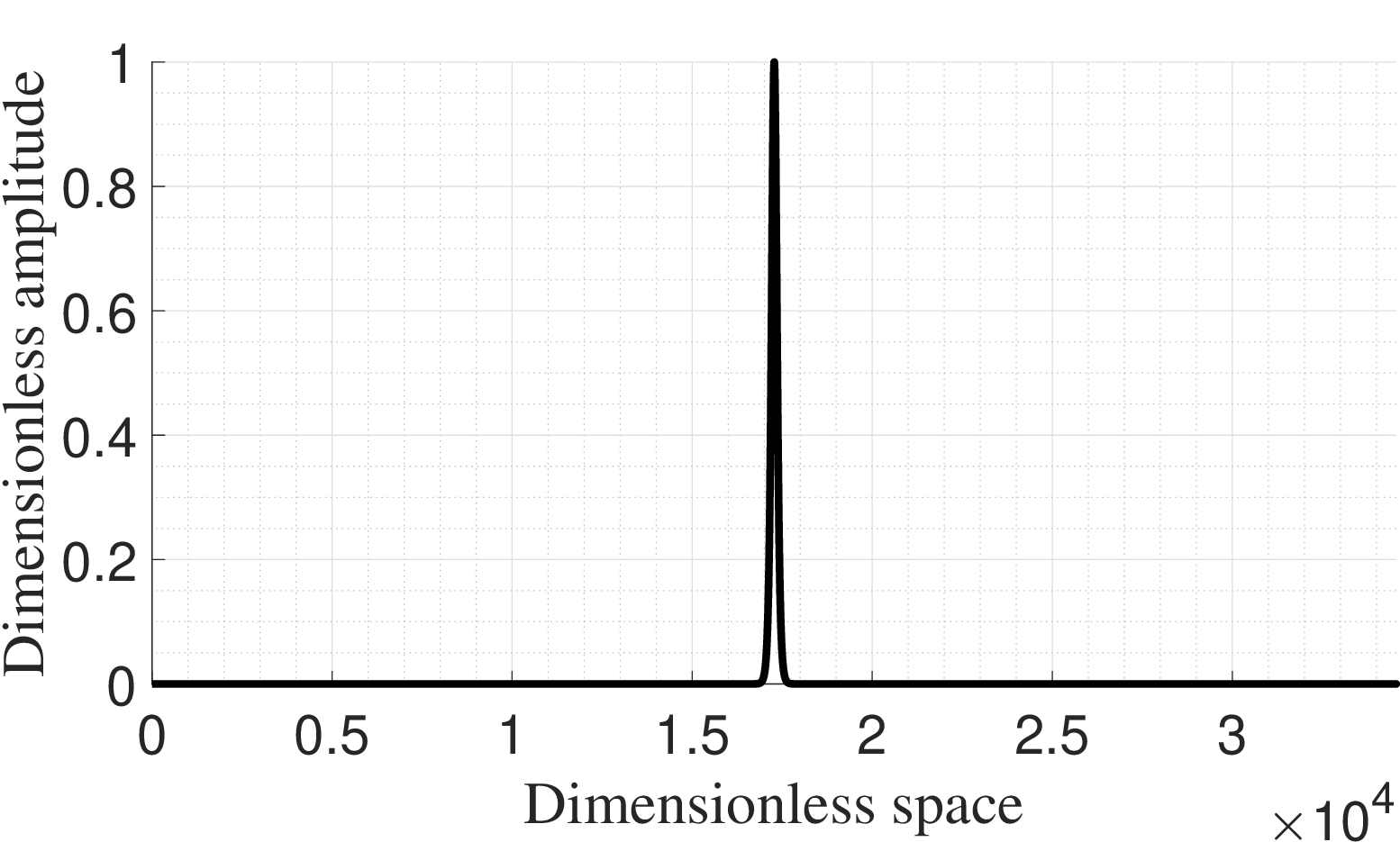}
    \includegraphics[width=0.49\textwidth]{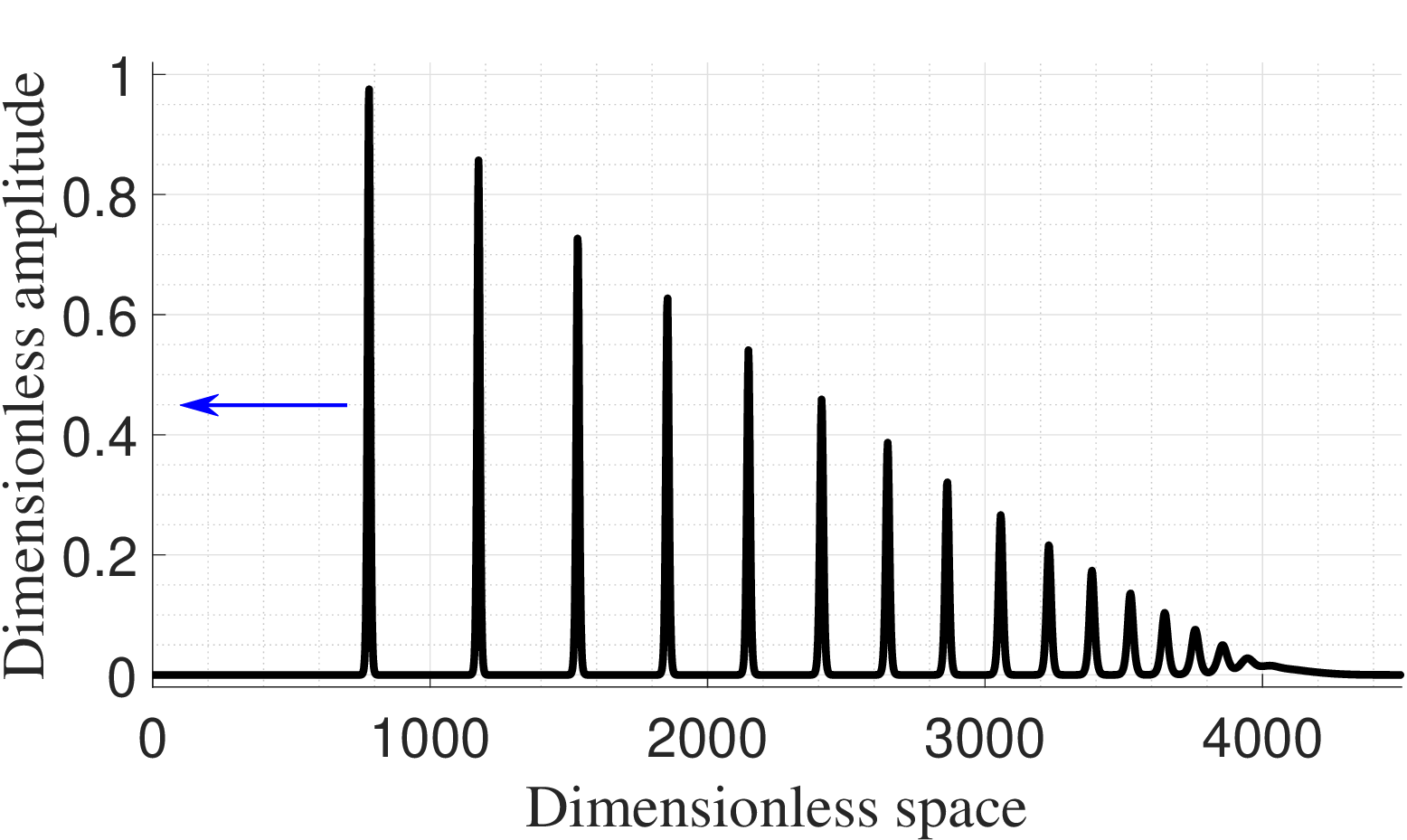}
    \caption{Formation of a soliton train in a Boussinesq-type equation from a localised initial disturbance.
See \cite{EngSalupTamm2011} for details (model equation~\eqref{Mindlin} with parameters
$\delta = 0.09, \, b = 0.7188, \, \beta = 56.0, \, \gamma = 9.3867, \, \mu = 1.1394, \, \lambda = 1.1470$).
Left panel: $\mathrm{sech}^2$-type initial condition, $U_0=1$, $B_0=0.01$.
Right panel: left-propagating train of solitons at $T=15700$. }\label{fig2uus}
\end{figure}

The mechanism of generating solitons has been studied since the seminal studies of Zabusky and Kruskal \cite{Zabusky1965} by many authors. These studies include the analysis of the one-wave equations like the Korteweg-deVries equation as well as the Boussinesq-type equations \cite{Remoissenet1999,PoS2006,Ablowitz2011,Raamat2021}, which are based on the modifications of the classical wave equation. Here, we demonstrate the formation of solitons according to a Boussinesq-type governing equation from a pulse-type initial excitation. The first example is the case of the microstructured solid where the governing equation in the one-dimensional case  (see eq.~\eqref{Mindlin} in the Appendix) includes terms related to nonlinear and dispersive effects \cite{EngSalupTamm2011}. Like in the case of the Korteweg-deVries equation  \cite{Zabusky1965}, a soliton train is generated (Fig.~\ref{fig2uus} right panel) from the localised initial condition (Fig.~\ref{fig2uus} left panel). The number of solitons in the train depends on the energy embedded in the initial excitation, and like in the classical Korteweg-deVries equation \cite{lauriandrus2009}, the larger the amplitude, the larger its velocity. 

\begin{figure}[p]
  \centering
    \includegraphics[width=0.49\textwidth]{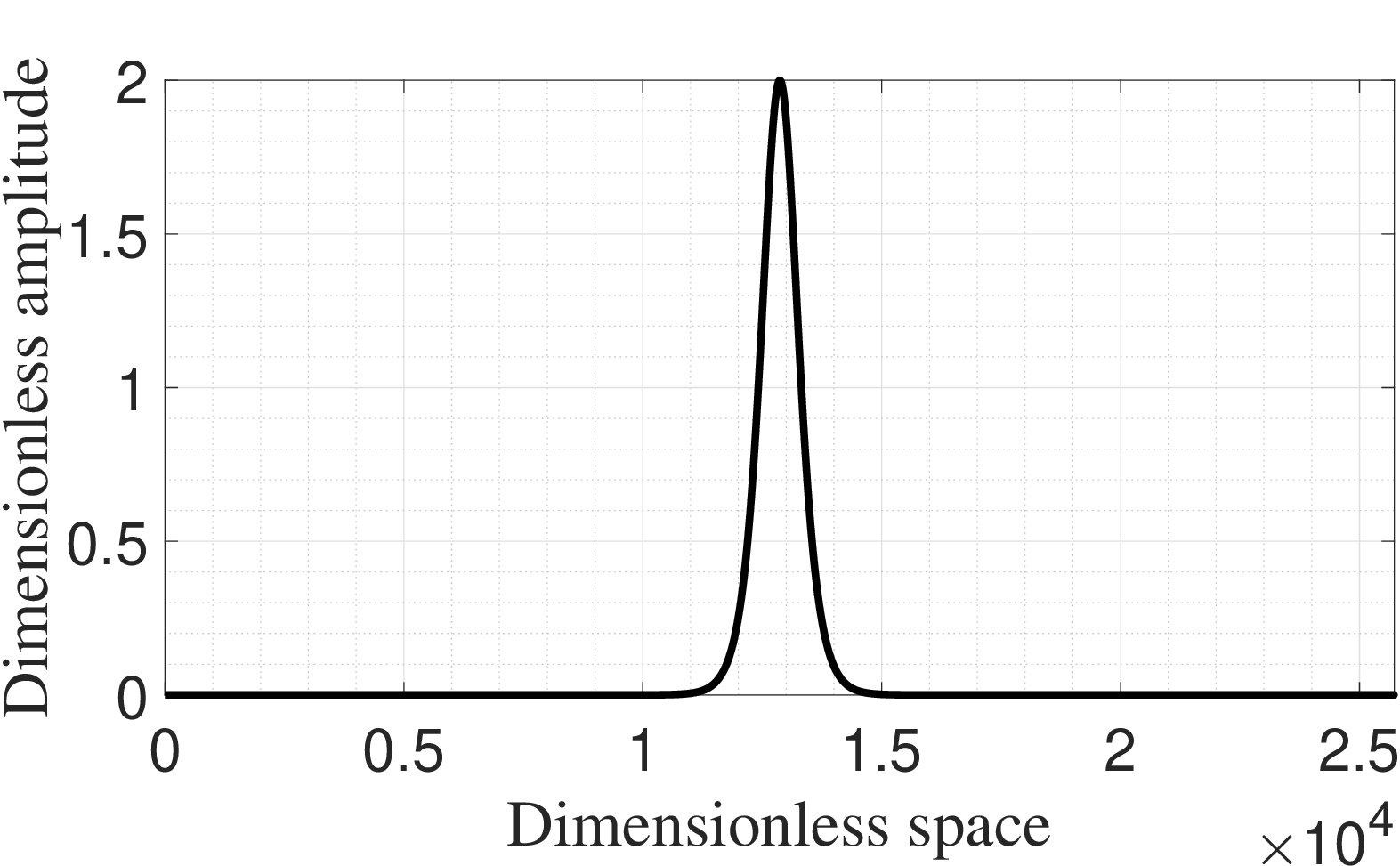}
    \includegraphics[width=0.49\textwidth]{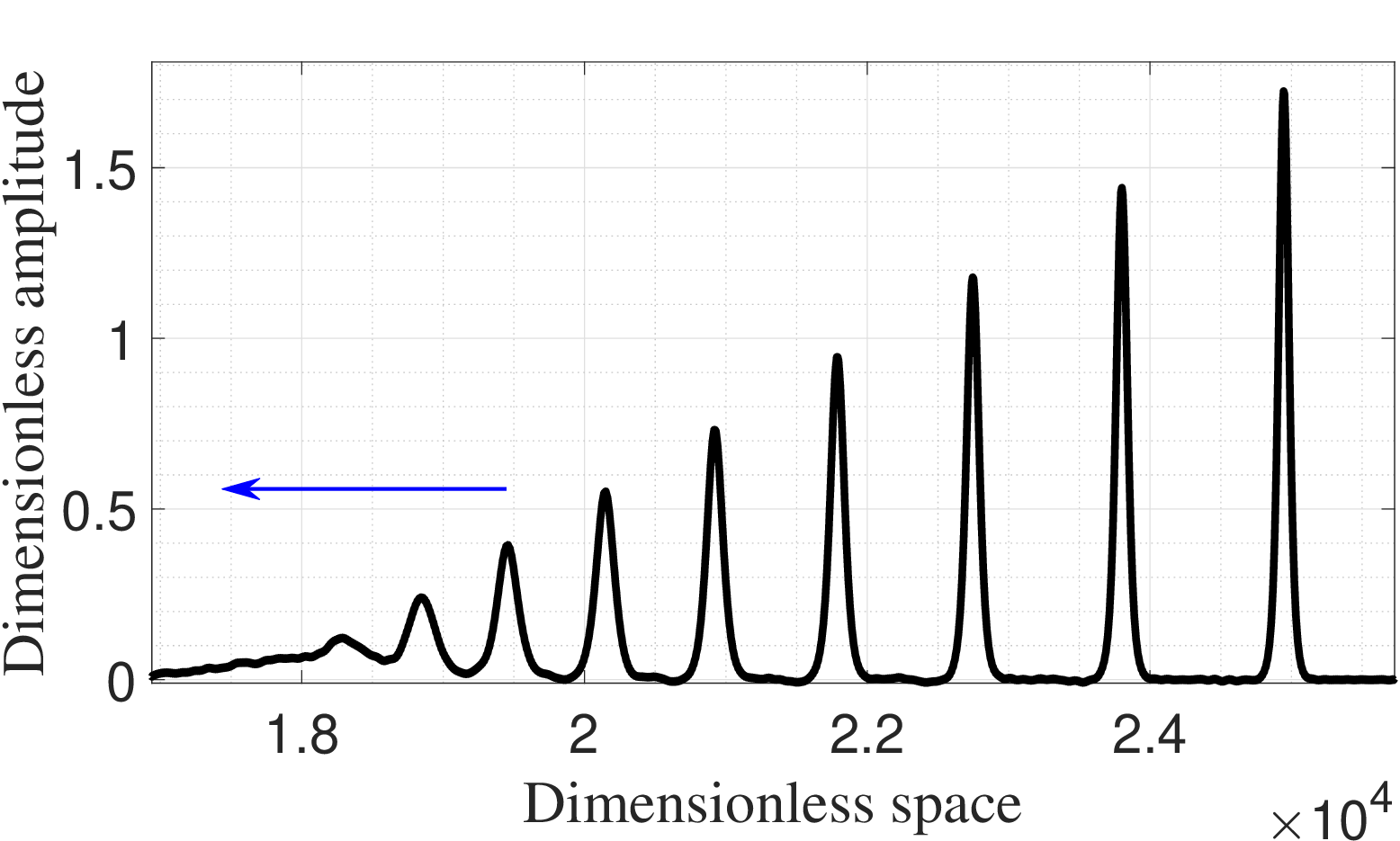}
    \caption{Improved Heimburg Jackson model \eqref{iHJ}, the formation of a soliton train from localised initial disturbance \cite{Tamm2015}.  Left panel -- initial profile. 
    Right panel - the formed soliton train at $T = 98 001$. Parameters $P = -0.2186, \, Q = 0.004230, \, H_1 = 72.14, \, H_2 = 1.000$ (see \cite{Tamm2015} for details).
}\label{fig3a}
\end{figure}
\begin{figure}[p]
  \centering
    \includegraphics[width=0.49\textwidth]{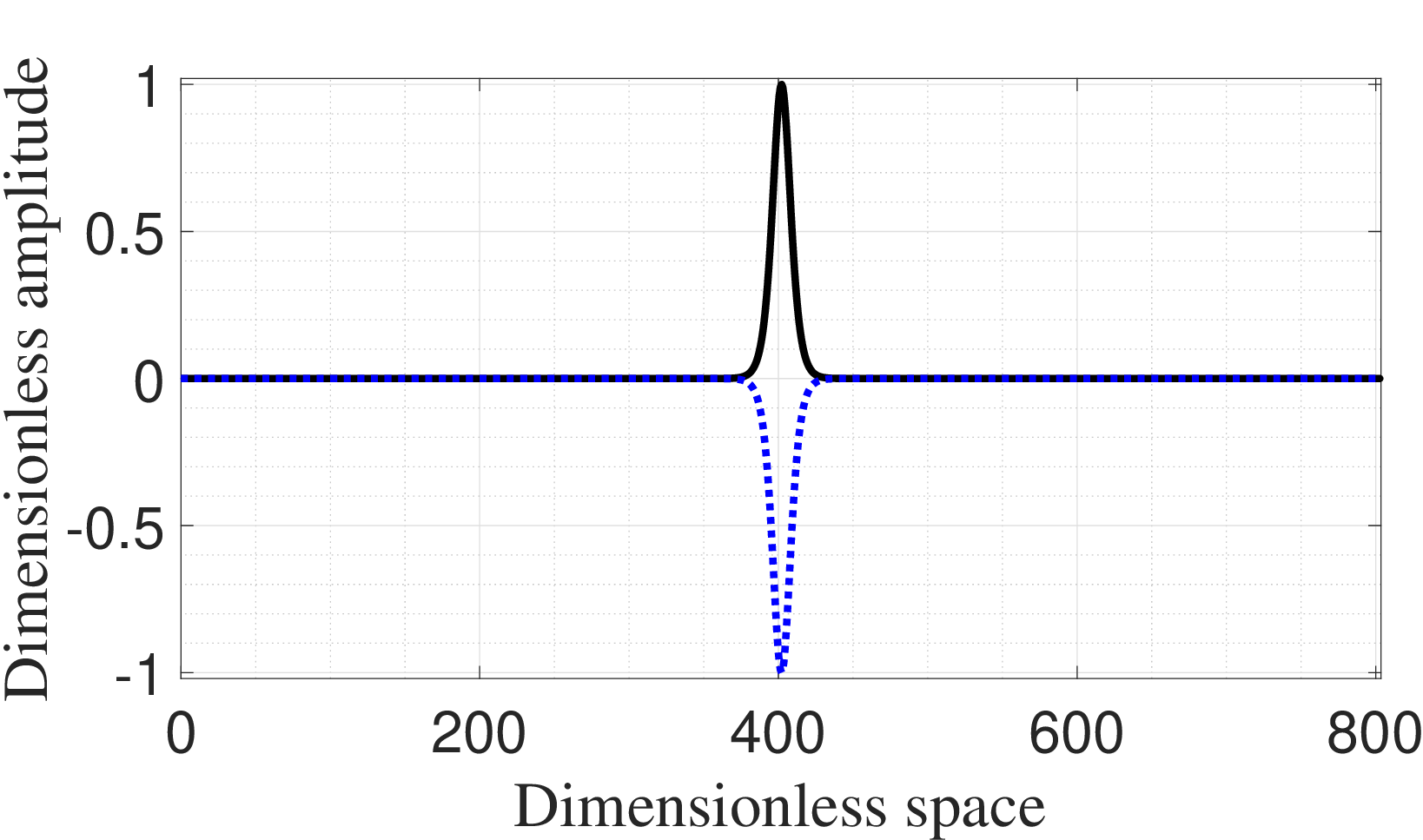}
    \includegraphics[width=0.47\textwidth]{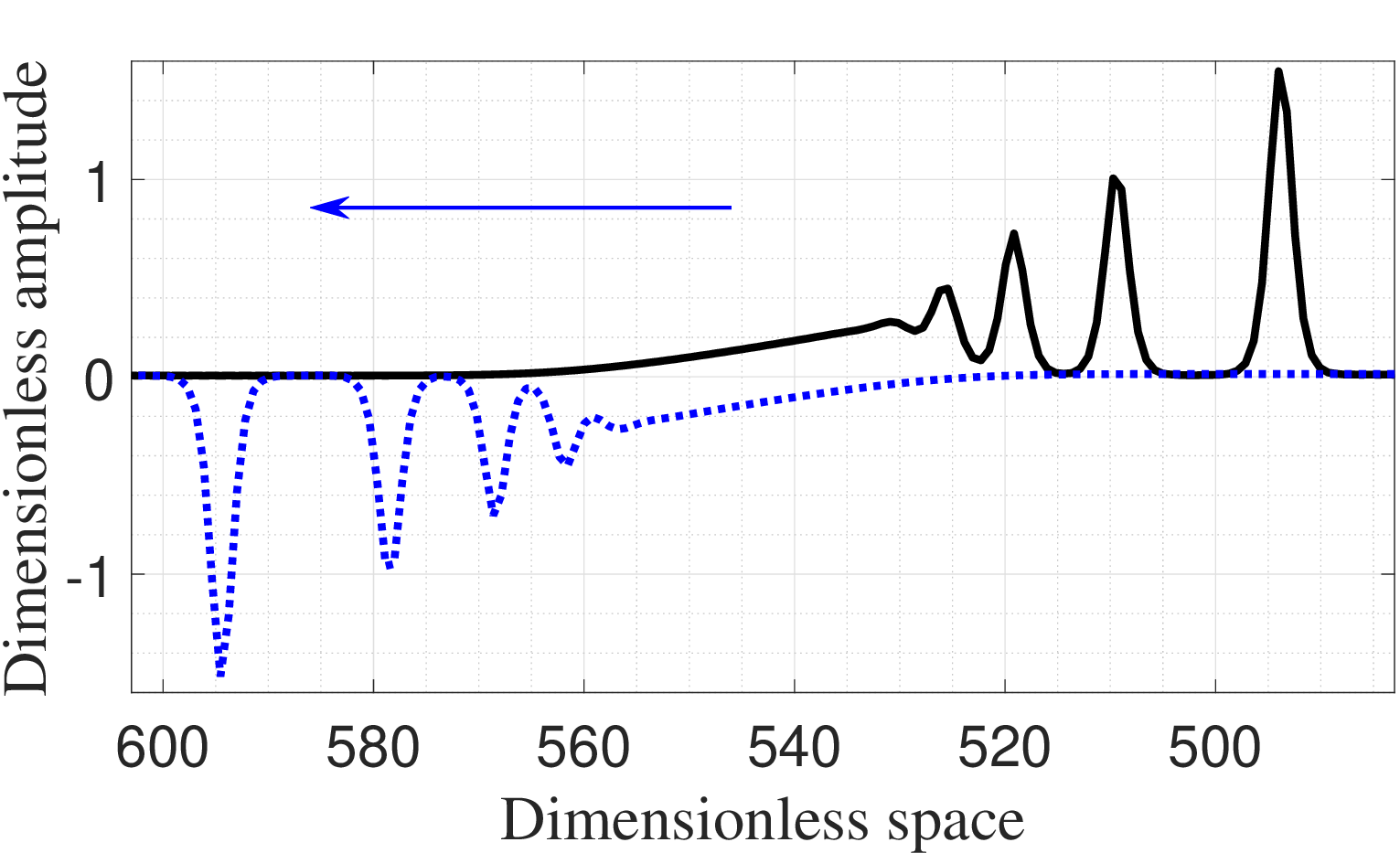}
    \caption{Improved Heimburg Jackson model \eqref{iHJ}, the formation of a soliton train from localised initial disturbance \cite{Engelbrecht2017}. Left panel - initial profiles with amplitudes 1 and -1. Right panel – positive and negative amplitude soliton trains (depending on the polarity of the initial condition) at $T = 1750$ (dimensionless). Dimensionless parameters $C_o = 1$, $P = -0.1$, $Q = 0.05$, $H_1 = 0.5$, $H_2 = 0.5$ (see \cite{Engelbrecht2017} for details).}\label{fig3b}
\end{figure}

\begin{figure}[p]
  \centering
    \includegraphics[width=0.49\textwidth]{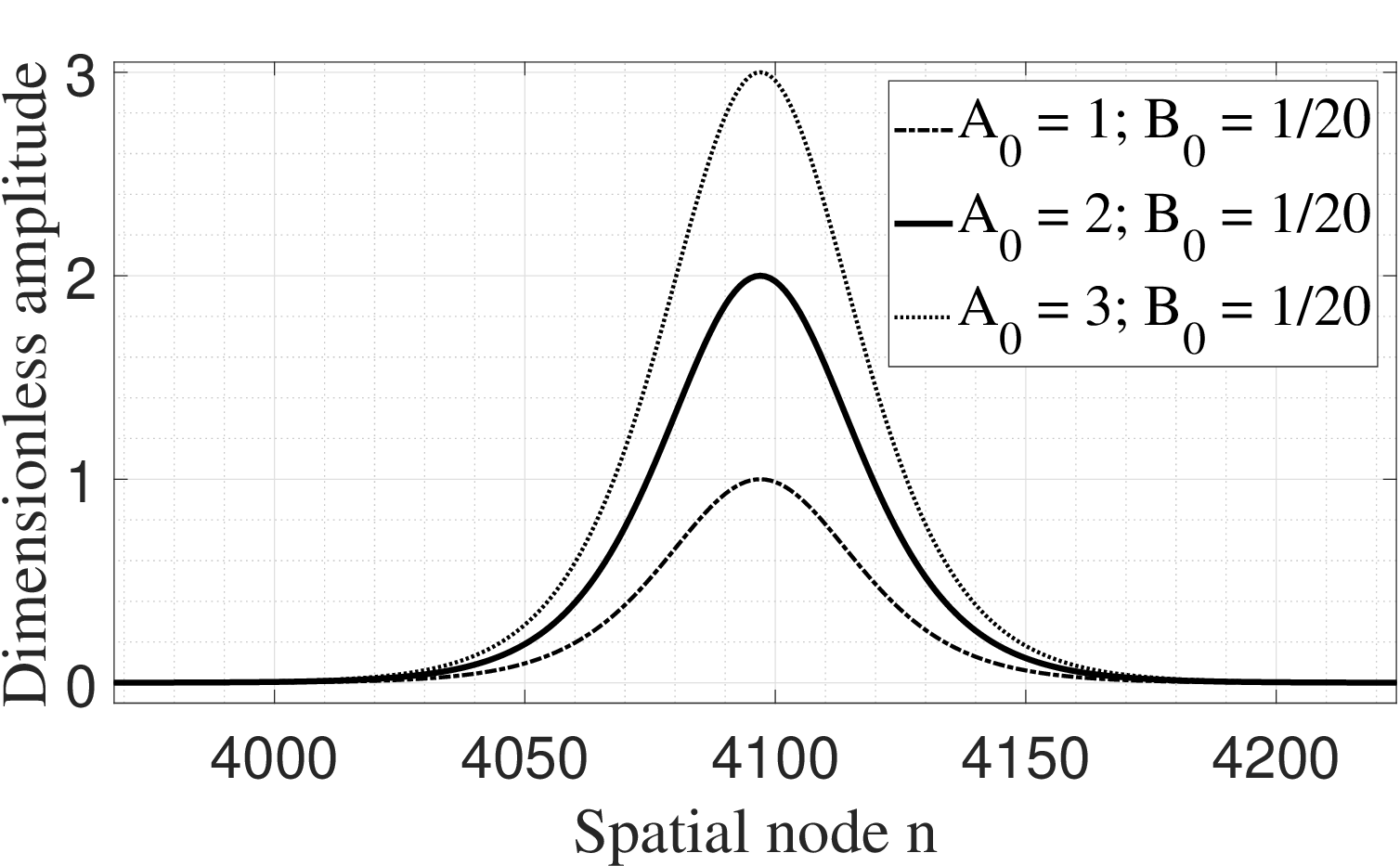}
    \includegraphics[width=0.49\textwidth]{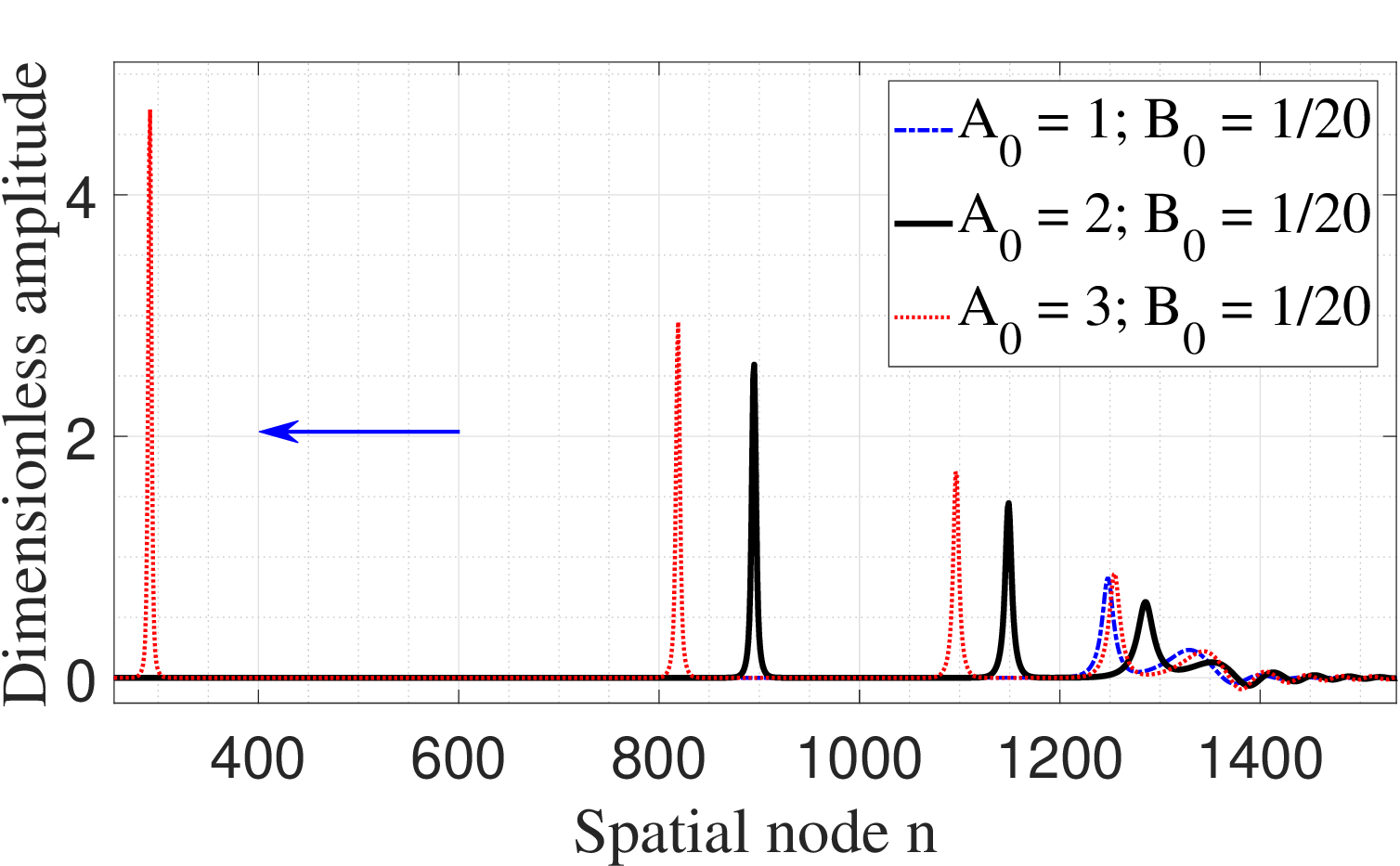}\\
    \includegraphics[width=0.49\textwidth]{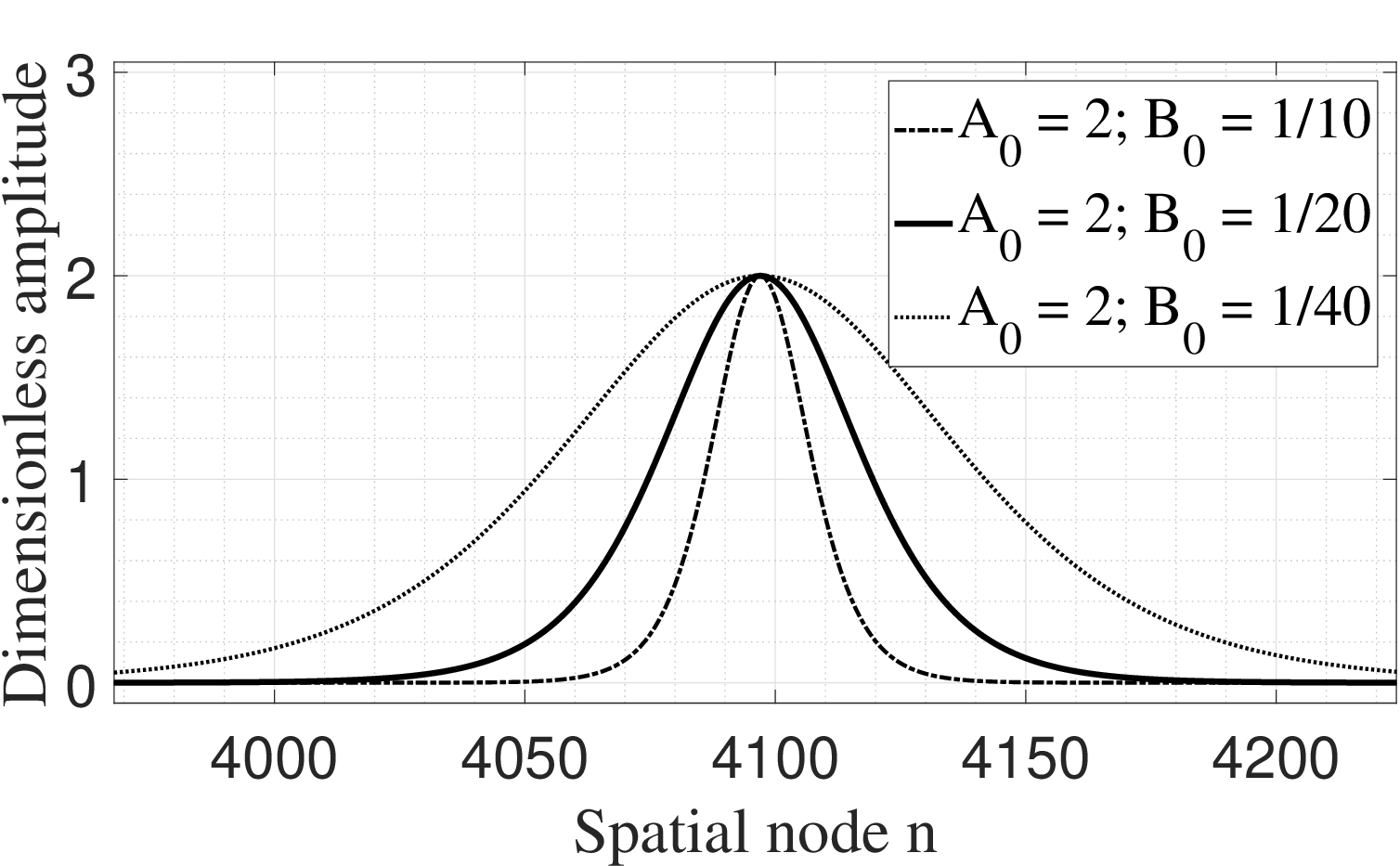}
    \includegraphics[width=0.49\textwidth]{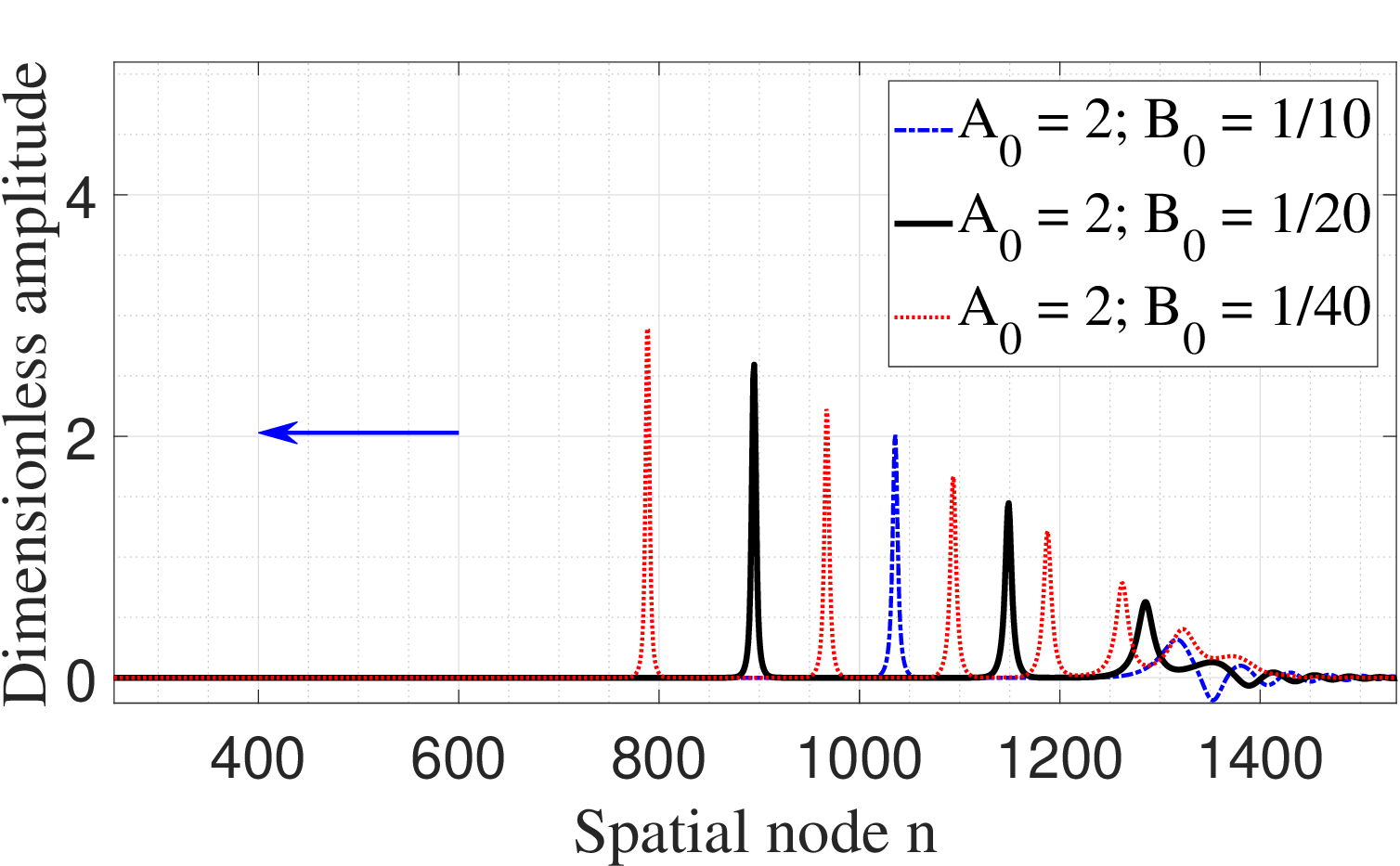}\\
    \caption{Formation of solitons from initial pulses with different widths and amplitudes for the iHJ equation. Parameters $P=-0.16$, $Q=0.8$, $H_1=2$, $H_2=3$. Space length $1024 \pi$ (dimensionless) and spatial grid $n=8192$. Time $T = 2198$ (dimensionless). Left - initial pulses (with zero initial velocity, splitting to two profiles propagating to the left and right); right - formed soliton trains and leftover oscillations, propagating to the left.  }\label{fig2a}
\end{figure}
\begin{figure}[p]
  \centering
    \includegraphics[width=0.32\textwidth]{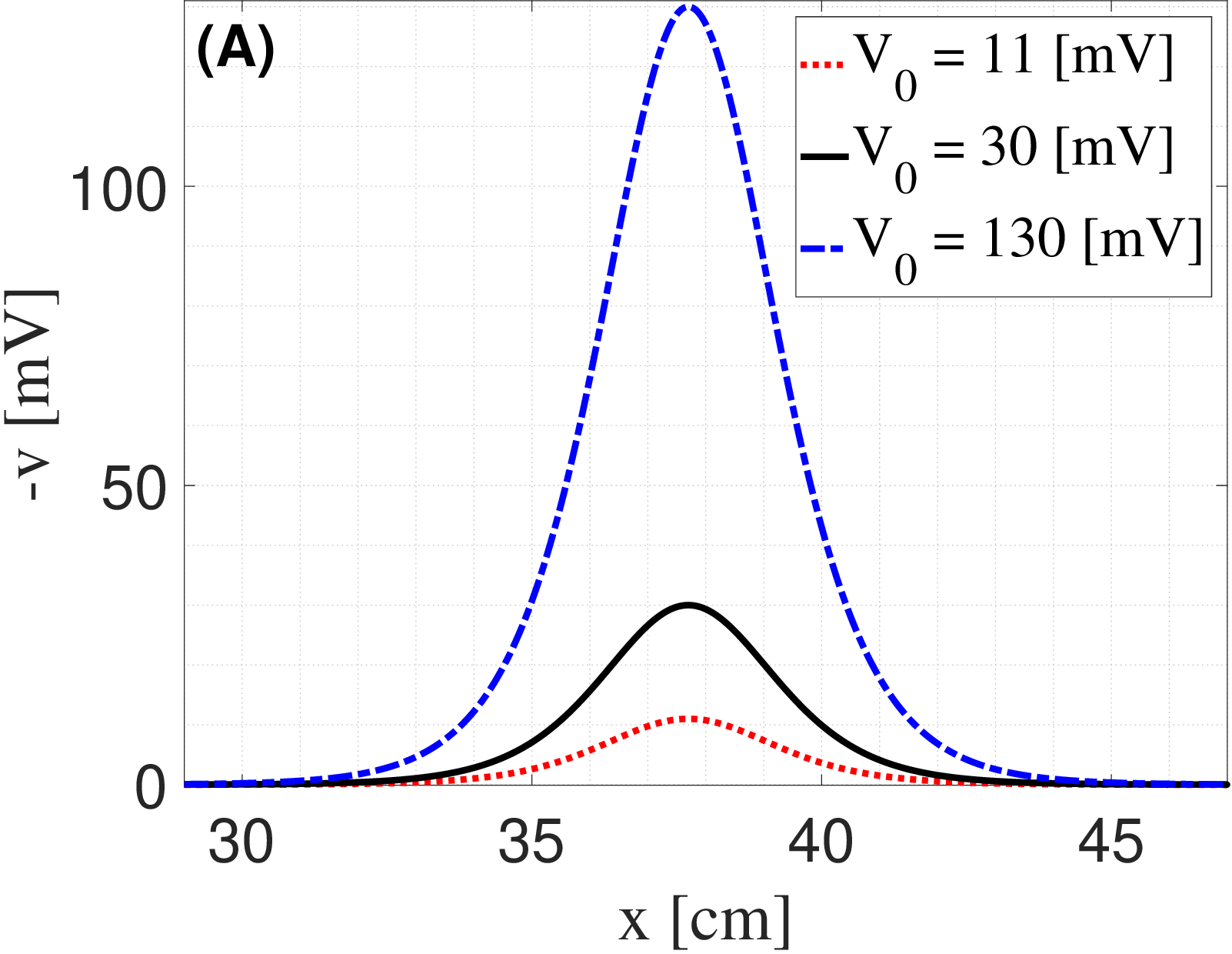}
    \includegraphics[width=0.327\textwidth]{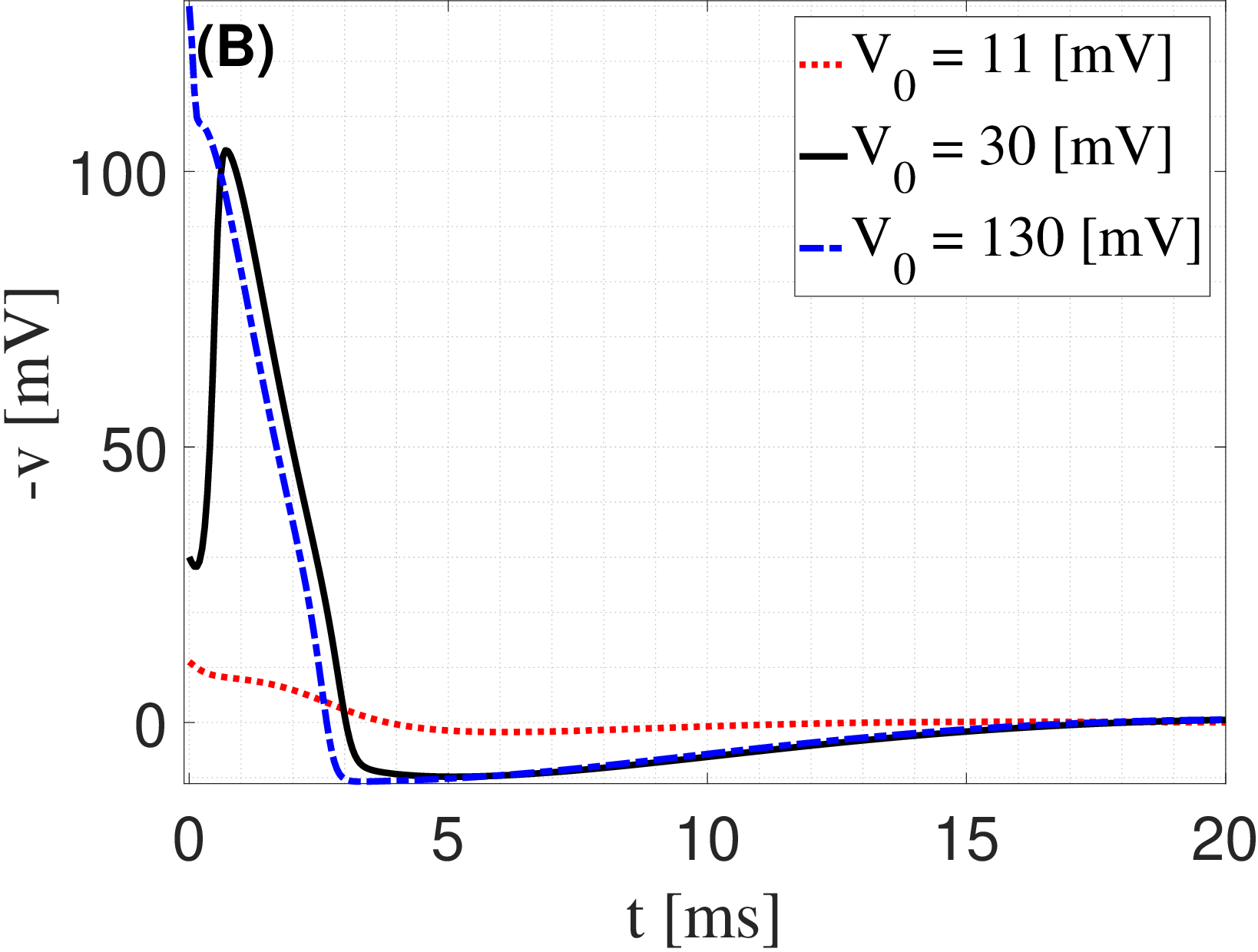}
    \includegraphics[width=0.32\textwidth]{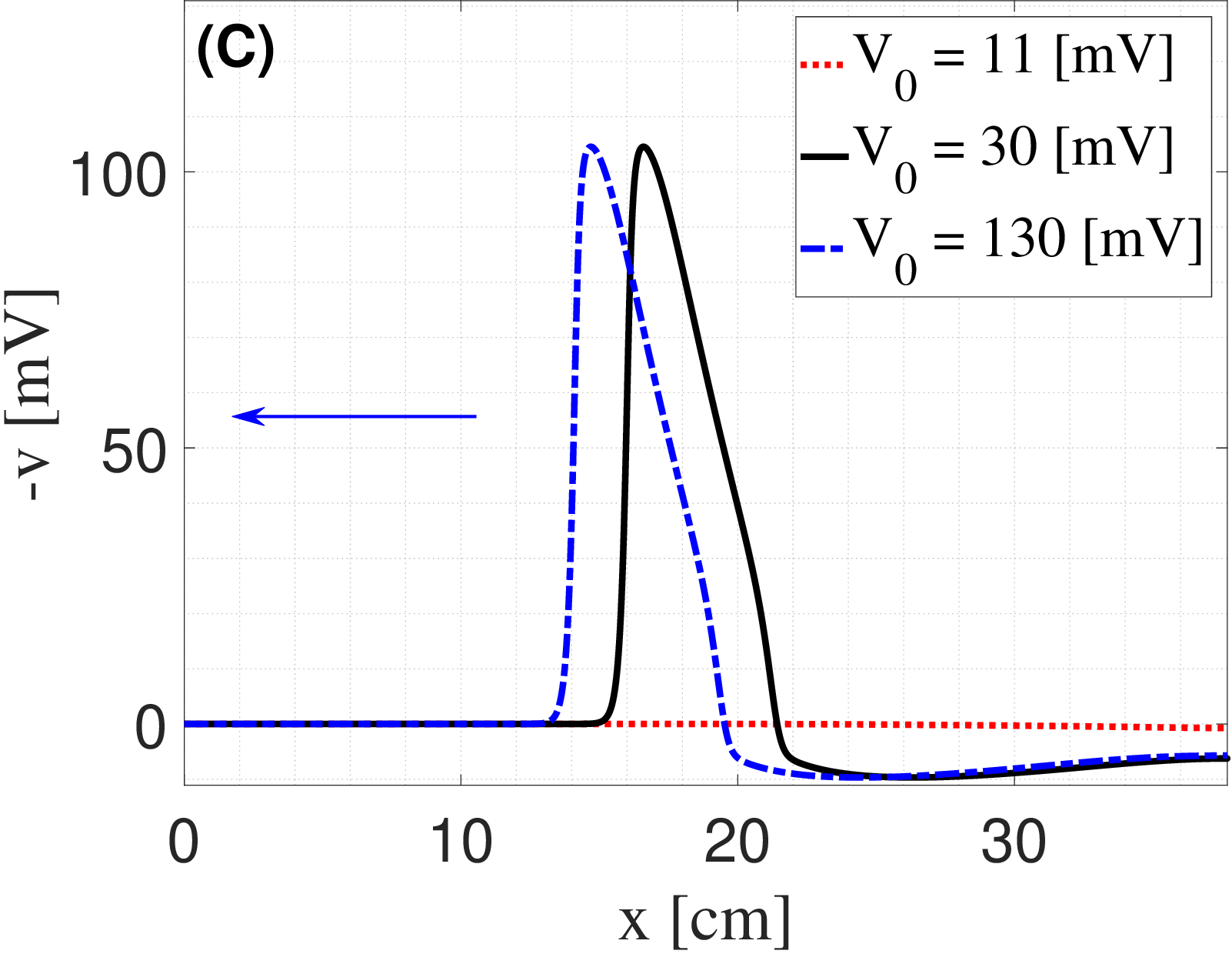}\\
    \includegraphics[width=0.32\textwidth]{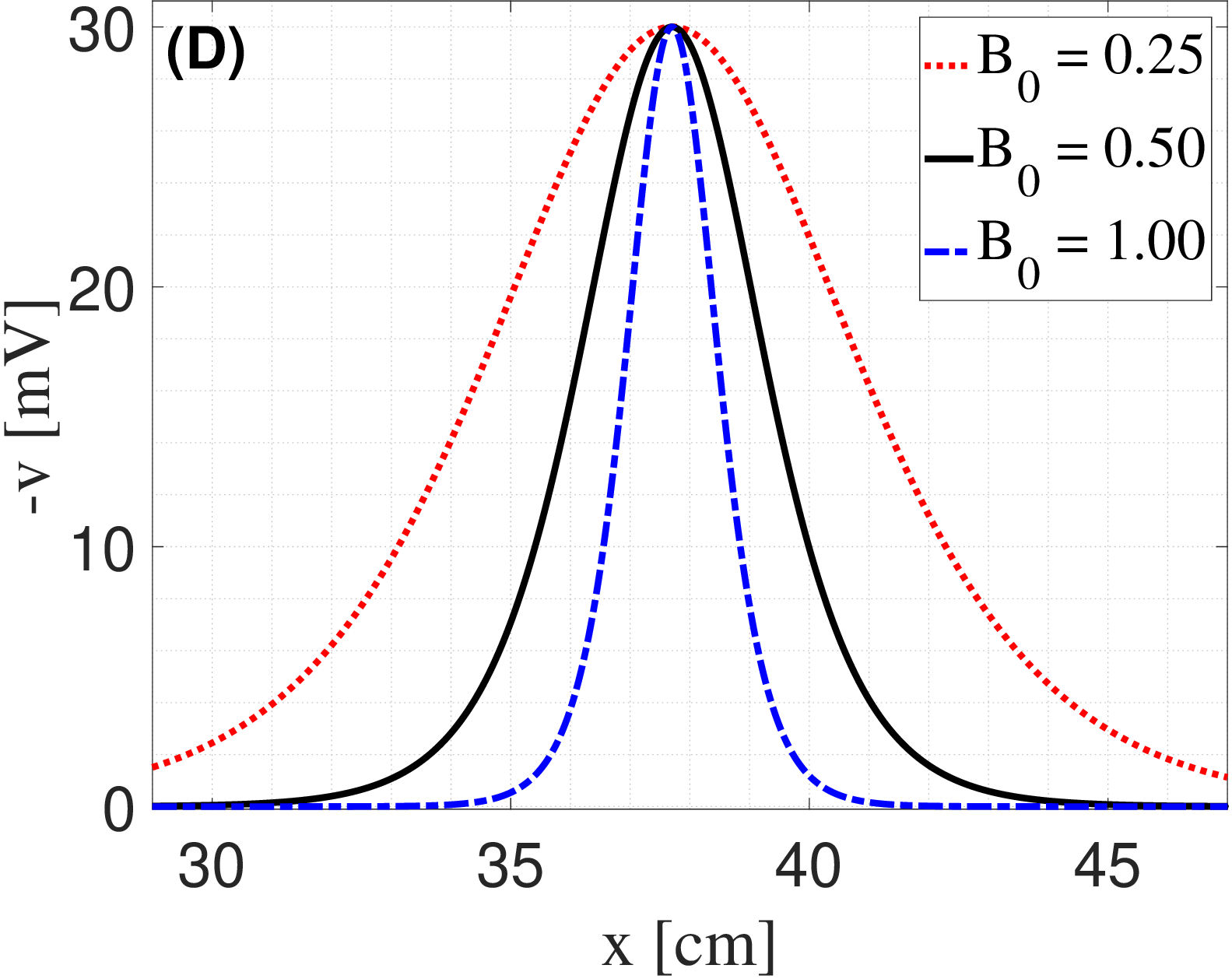}
    \includegraphics[width=0.327\textwidth]{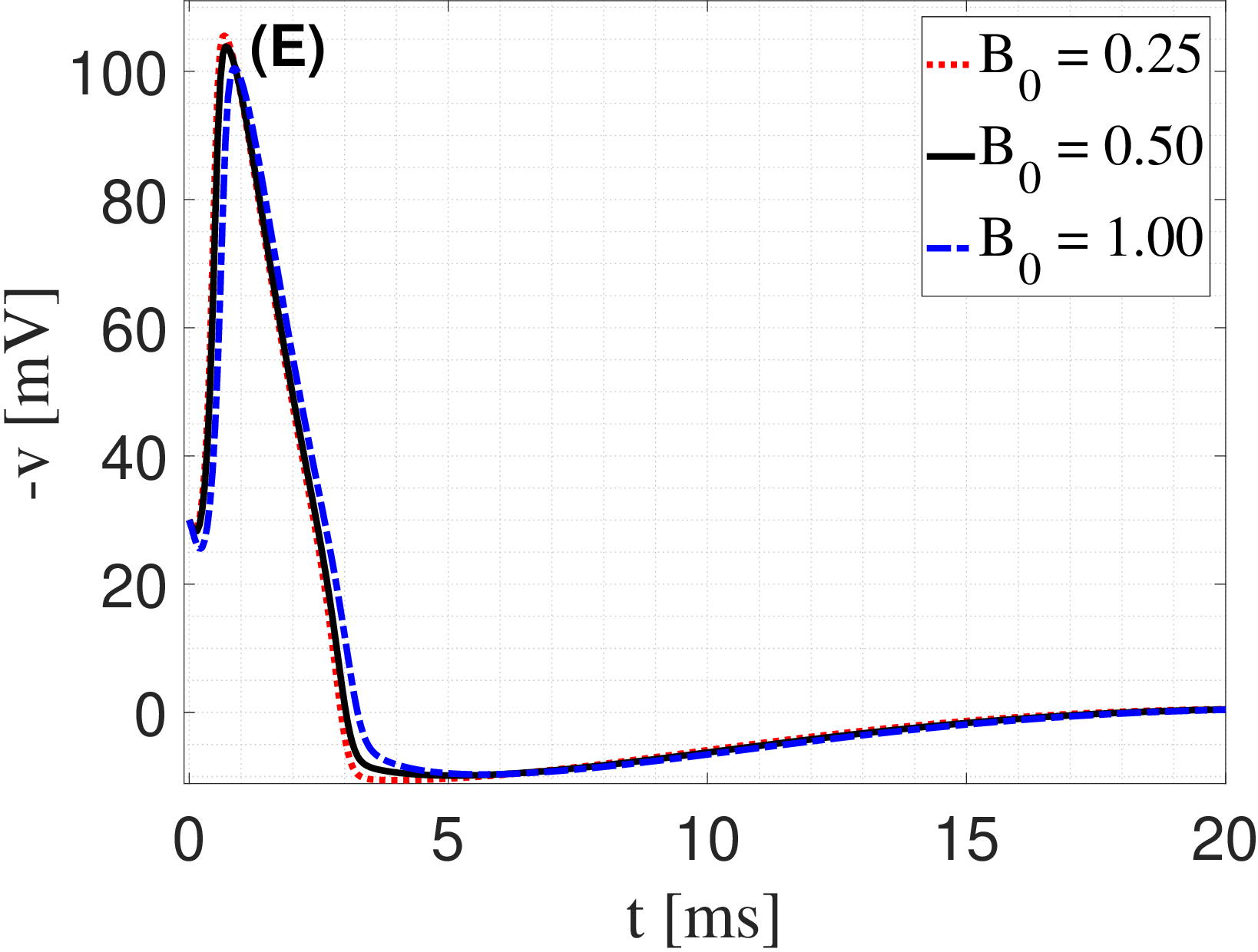}
    \includegraphics[width=0.32\textwidth]{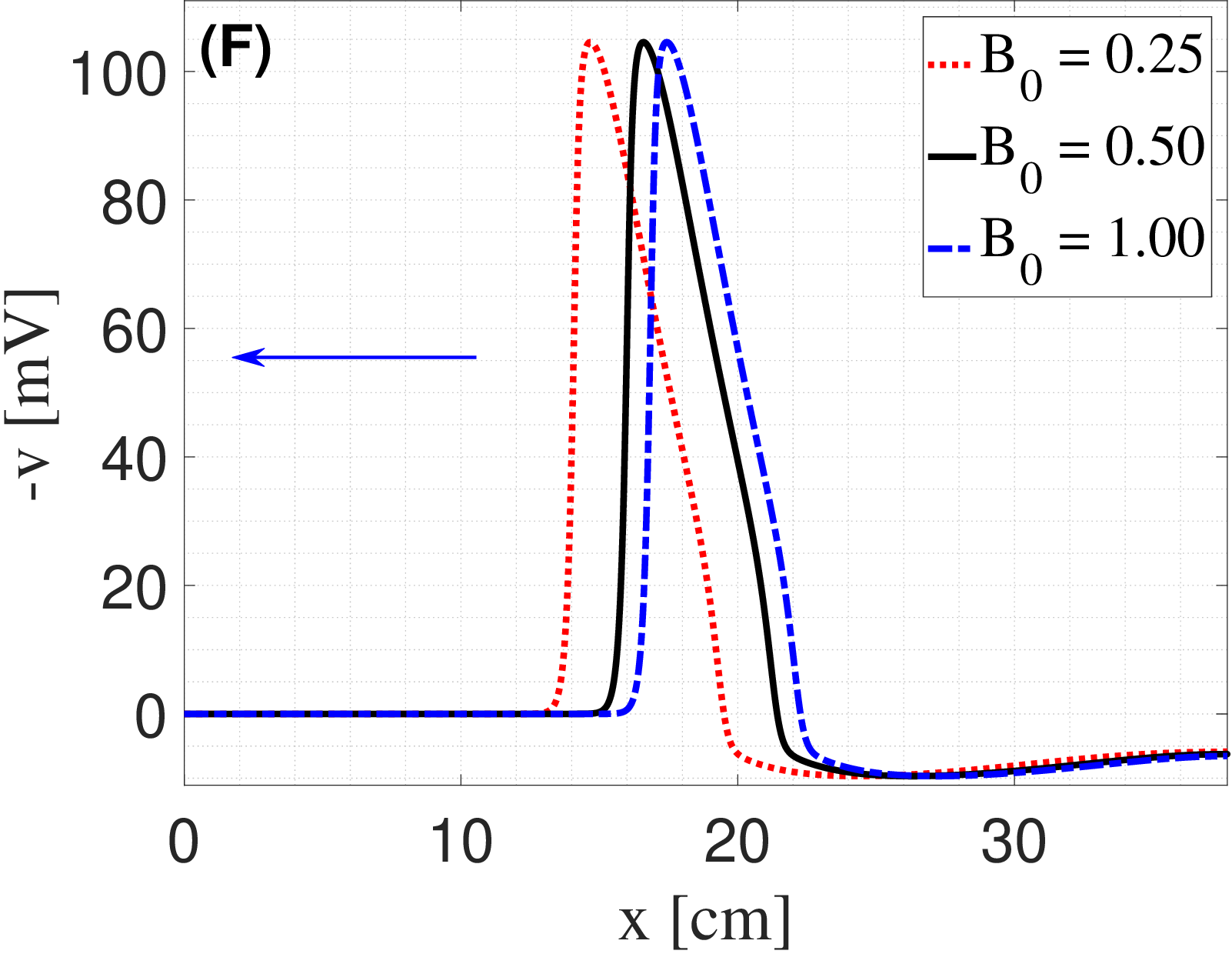}
    \caption{Formation of the action potential from an initial pulse with different initial amplitudes \cite{Tamm2025}. Threshold value $V_0 \approx 12 \mathrm{[mV]}$. In the top panel \textbf{(A)} is initial excitation profiles in the middle of the spatial period at $t=0 \mathrm{[ms]}$, panel \textbf{(B)} is AP signals in time at $x=37.67 \mathrm{[cm]}$ (peak location of initial pulse), panel \textbf{(C)} is AP signals in space ($x=0 \ldots 37.67 \mathrm{[cm]}$) at $t=10 \mathrm{[ms]}$.  In the bottom panel \textbf{(D)} is initial excitation profiles in the middle of the spatial period at $t=0 \mathrm{[ms]}$, panel \textbf{(E)} is AP signals in time at $x=37.67 \mathrm{[cm]}$ (peak location of initial pulse), panel \textbf{(F)} is AP signal in spaces ($x=0 \ldots 37.67 \mathrm{[cm]}$) at $t=10 \mathrm{[ms]}$.}\label{fig2b}
\end{figure}
\begin{figure}[p]
  \centering
    \includegraphics[width=0.92\textwidth]{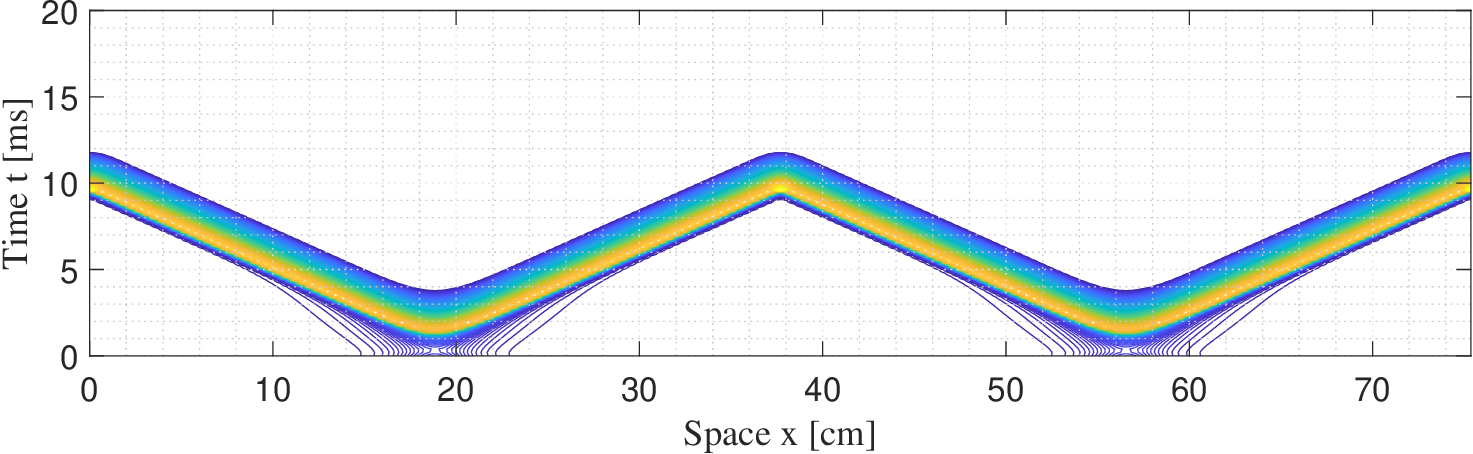}\\
    \includegraphics[width=0.92\textwidth]{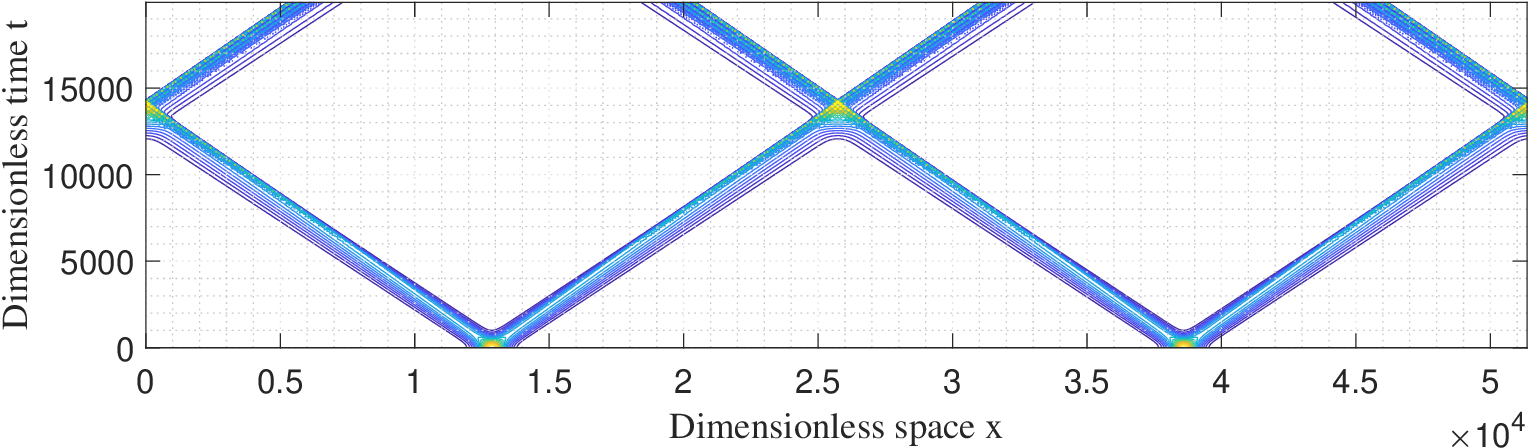}
    \caption{Contour plots (isolines) demonstrating the interaction of solitary pulses (bottom panel, model eq.~\eqref{iHJ}) versus interaction of counter-propagating AP signals (top panel, model eqs.~\eqref{LIB11} and \eqref{LIB21}). In both cases, periodic boundary conditions were used in space. Note how the AP signal is annihilated upon head-on collision, while the mechanical wave in the cell membrane continues to propagate after the head-on collision.}\label{fig5}
\end{figure}

In Fig.~\ref{fig3a}, one can see a formation of a soliton train for the improved Heimburg-Jackson (iHJ) model \eqref{iHJ} with the parameters as they are measured for the biomembrane \cite{Tamm2015,Heimburg2005}. One should note that with these parameters, the lower amplitude solitons travel faster and the higher amplitude ones slower \cite{Vargas2018,Heimburg2005}. However, by changing the physiological constraints and allowing parameter combinations beyond physiologically plausible ones, the iHJ model can support a wider range of behaviours, like, for example, a scenario where the polarity of the forming soliton train depends on the polarity of the initial excitation in a certain narrow band of parameters (see Fig.~\ref{fig3b}) \cite{Engelbrecht2017}. 

One should note that the formation of solitons and soliton trains from the localised pulse can take some time, as an example, the formed soliton train depicted in Fig.~\ref{fig3a} right panel corresponds to roughly 110 milliseconds of time and the propagation distance of roughly two meters (if axon diameter is 20 {\textmu}m), which is significantly more than the typical dimensions of nerve axons. It should be stressed that even if the model supports interesting solutions, one should be careful when interpreting the results in the biological context, as there can exist additional physiological constraints not explicitly included in the model during the model derivation. Even before considering the question if all assumptions made during derivation, like taking a conservative system (no dissipation or energy inflow), are justified for describing processes in the cell membrane over such propagation distances and evolution times.

When a soliton train forms from an arbitrary initial excitation, what is happening depends on the energy inserted into the system. It must be stressed here that typically the nonlinear dispersive model supporting solitonic solutions is conservative \cite{Ablowitz2011}.
Over a sufficiently long time, a soliton train is formed, and the number of solitons (if the parameters of the system allow solitonic solutions) depends upon the energy of the excitation. The ``leftover'' energy will form an oscillatory ``tail''. This is demonstrated in Fig.~\ref{fig2a} for different initial excitation amplitudes (top panels) and different initial excitation widths (bottom panels) in the case of the iHJ model \eqref{iHJ} in which with the used parameters higher amplitude solitons are faster.
 
In contrast, all the action potential models that are commonly considered as physiologically realistic by the scientific community (usually some version of the Hodgkin-Huxley type model) describe a threshold mechanism. This means that any excitation below a certain threshold value dissipates rapidly and any excitation above the threshold generates the propagating action potential signal. This is demonstrated in Fig.~\ref{fig2b} for the so-called Lieberstein model (a version of Hodgkin-Huxley model where inductance has not been neglected) \cite{Tamm2025,Tamm2026,Lieberstein1967} where excitation below threshold (red dotted line, panels (A), (B), (C)) does not generate the AP and excitation above the threshold generates basically the same AP signal regardless of the input signal amplitude (solid and dash-dotted lines, panels (A), (B), (C)). For the sake of completeness, it should be noted that the width of the initial excitation does not significantly affect the formation of the AP signal (Fig.~\ref{fig2b} panels (D), (E), (F)). The initial pulse width affects only the time needed for the emergence of the AP. 
Another notable difference is that solitons typically interact elastically with a small phase shift due to the nonlinear effects (Fig.~\ref{fig5} bottom panel) while the action potentials annihilate during a head-on collision (Fig.~\ref{fig5} top panel). The interaction of solitons is analysed in \cite{Ostrovsky2022} in detail including also non-integrable systems in which the collision of solitons is characterised by accompanying radiation (like in iHJ model \eqref{iHJ} \cite{Engelbrecht2017} and MEP model \eqref{Mindlin} \cite{EngSalupTamm2011}). 

\clearpage

\section{Final remarks}\label{Final}

Although Scott \cite{Scott1999} has already clearly explained the differences between APs and solitons, there seems to be a need to stress these differences again. The notion of a soliton is indeed attractive for describing solitary waves, but the conditions under which a wave can be called a soliton are clearly fixed and agreed upon within the framework of mathematical physics. The definitions (see Section \ref{Definitions}) and examples (see Section \ref{Propagation}) demonstrate clearly the differences. However, as shown in many experiments, the propagation of an AP is accompanied by mechanical deformation of the biomembrane. It is proposed by Heimburg and Jackson \cite{Heimburg2005} that the longitudinal mechanical wave (LW) in the biomembrane can be described by a Boussinesq-type equation, which has soliton-type solutions. The family of these solutions is indeed rich \cite{Engelbrecht2017}, but one should pay attention to the emergence of solitons from an arbitrary excitation. The timescale of emerging solitons (time-scale of roughly hundred milliseconds and propagation distance in meters for the first solitons to separate from the arbitrary initial disturbance \cite{Engelbrecht2017}) in this case exceeds the timescale of the existence of APs. The coupling of an AP with the accompanying LW is analysed by Engelbrecht et al. \cite{Raamat2021} based on the hypothesis that all accompanying physical effects are caused by the changes in the electrical field (i.e. by an AP) in the nerve fibre. Then the LW is generated by a driving force in the governing Boussinesq-type equation, resulting in an asymmetric solitary wave which propagates in the phase with the AP. 

Besides the ion mechanism proposed by Hodgkin and Huxley \cite{Hodgkin1952}, the role of density pulse in the biomembrane proposed by Heimburg and Jackson \cite{Heimburg2005} has generated a discussion about the thermodynamics of APs. As stated by Drukarch and Wilhelmus \cite{Drukarch2026}, this discussion may open the interdisciplinary perspective leading to a comprehensive explanation of neuronal excitability. It should be noted that while the HJ and iHJ models were a reasonable starting point for modelling mechanical processes in cell membrane, van der Waals type models like, for example, \cite{Mussel2019a,Das2025} appear to be capable of capturing some phenomena associated with propagating action potentials, like annihilation during head on collision better. 
However, the terminology of mathematical physics should be used properly. The notions of ``soliton'' and ``solitary wave'' are not synonyms, and solitons are not appropriate for describing nerve signals.
\appendix

\section{Model equations, initial and boundary conditions and numerical method}\label{}
A class of nonlinear partial differential equations forms the so called Boussinesq paradigm \cite{Christov2007} ``$\ldots$ this set of equations which contains simultaneously the following ingredients: (i) bi-directionality of the wave solutions (propagation to the left and to the right; presence of a d’Alembertian operator); (ii) nonlinearity of any order; and (iii) dispersion of any order, the latter resulting in the presence of combined space and time derivatives of the fourth order at least''. One member of this family of equations is the Mindlin-Engelbrecht-Pastrone (MEP) model \cite{EngSalupTamm2011,TammSalupereA2010} 
\begin{equation}\label{Mindlin}
V_{TT} - bV_{XX} - \frac{\mu}{2} \left(V^2\right)_{XX} = \delta \left( \beta V_{TT} - \gamma V_{XX} + \frac{\lambda \sqrt{\delta}}{2} \left(V_X\right)_X^2 \right)_{XX},
\end{equation}
\begin{displaymath}
\text{in eq.~\eqref{Mindlin}} \quad b = 1 - \frac{D^2}{AB}, \, \mu = \frac{NU_{0}}{AL_{0}}, \, \beta = \frac{ID^2}{\rho l_{0}^{2} B^2}, \, \gamma = \frac{CD^2}{AB^2 l_{0}^{2}}, \, \lambda = \frac{D^3 M U_{0}}{AB^3 l_{0}^{3} L_{0}} ,
\end{displaymath}
where $V$ is deformation, $A,B,C,D$ are material coefficients, $N,M$ are nonlinear coefficients, $\rho$ is density, $I$ is microstrcture inertia related parameter, $l_{0}$ is the characteristic scale of microstructure, $U_0$ is the amplitude of the initial excitation, $L_0$ is  the wavelength of the initial excitation and $\delta$ is a scale factor. Subscript $X$ denotes partial derivative by space and $T$ partial derivative by time. Like is common for Boussinesq-type equations, this too supports soliton-type solutions in some parameter ranges (see Fig.~\ref{fig2uus}).

The improved Heimburg Jackson (iHJ) model \cite{Heimburg2005,HJ2007,Peets2015,Raamat2021}
\begin{equation}\label{iHJ}
U_{TT} = C_{0}^{2} U_{XX} + P U U_{XX} + Q U^2 U_{XX} + P U_{X}^{2} + 2 Q U U_{X}^{2} -  H_1 U_{XXXX} + H_2 U_{XXTT},
\end{equation}
where $C_{0}$ is the velocity of unperturbed state in lipid bi-layer, $P, Q$ are the nonlinear coefficients, $H_1, H_2$ are the dispersion coefficients in the dimensionless case (see \cite{Raamat2021} for details).  A $\textrm{sech}^{2}$-type localised initial condition with initial amplitudes $A_0$ and width $B_0$ are used, and we make use of the periodic boundary conditions
\begin{equation}
 U(X,0) = A_{0} \textrm{sech}^2 B_{0} X, \; U_T(X,0) = 0, \; U(X,T) = U (X + 2 K m \pi,T), \; m = 1,2,\ldots
\end{equation}
where $K$ is the number of  $2\pi$ sections in the spatial period. Please note the zero initial velocity, which means that the initial profile splits into two counter propagating profiles with half the initial amplitude, which then start evolving in time due to the nonlinear and dispersive effects.

Equation \eqref{iHJ} has analytical soliton-type solution \cite{Engelbrecht2017} with suitable constraints on parameters in the form 
\begin{equation}\label{HJsoliton}
U(\xi) = \frac{6(c^2 - C_{0}^{2})}{P \left( 1 + \sqrt{1 + \frac{6Q(c^2 - C_{0}^{2})}{P^2}} \cosh \left[ \xi \sqrt{\frac{C_{0}^{2} - c^2}{H_1 - H_2 c^2}}\, \right] \right)},
\end{equation}
where $c$ is the velocity of the soliton and $U$ is the amplitude. For the sake of completeness equation \eqref{Mindlin} can have soliton-type analytical solutions \cite{PeetsTammEngelbrecht2017} (in moving frame of reference $\xi$) with suitable constraints on parameters:
\begin{equation}
V(\xi) = A_0 \operatorname{sech}^2 \left[ \frac{1}{2} \kappa \xi \right] = \frac{2 A_0}{1+\operatorname{cosh} \left[\kappa \xi\right]}, \quad \text{where} \quad A_0 = \frac{3(c^2 - b)}{\mu}, \; \kappa = \sqrt{\frac{c^2 - b}{\delta(\beta c^2 - \gamma)}} \; \text{and} \; \xi = X - c \cdot T.
\end{equation}

The Lieberstein model \cite{Tamm2025} is used in a form where action potnetial \eqref{LIB11} and ionic currents across the membrane \eqref{LIB21} are written as individual equations
\begin{equation} \label{LIB11}
\left(C_m 2 \pi a\right)\frac{\partial v}{\partial t} + \frac{\partial i_a}{\partial x}
+ 2 \pi a \cdot \left[\hat{g}_K n^4 (v-V_K) + \hat{g}_{Na} m^3 h (v-V_{Na}) + \hat{g}_l (v-V_l)  \right] = 0,
\end{equation}
\begin{equation} \label{LIB21}
\frac{L}{\pi a^2}\frac{\partial i_a}{\partial t} + \frac{\partial v}{\partial x} + r i_a = 0.
\end{equation}
Equations~\eqref{LIB11} and \eqref{LIB21} are solved numerically with the pseudospectral method \cite{Tamm2025}, although we have recently also used a finite volumes based numerical scheme which yields similar results \cite{zeeshan2026}. Here $x$ is space (length) and $t$ is time; $v$ is the action potential, $i_a$ is the line axon current (along the axon) and $i$ is the membrane current per unit length; $a$ is the radius of the axon, $r$ is the axon resistance per unit length, $L=22.2 \, [mH \cdot cm]$ is the axon specific self-inductance and $C_m$ is the membrane capacity per unit area. Coefficients $n,m,h,\hat{g}_K,\hat{g}_{Na},\hat{g}_l,V_K,V_{Na},V_l$ are taken the same as in \cite{Hodgkin1952}.
\begin{center}
\begin{tabular}{ | c | c | c | }
\hline
\multicolumn{3}{|c|}{Table 1: Parameters for eqs.~\eqref{LIB11} and \eqref{LIB21} (see also \cite{Hodgkin1952} {and } \cite{Lieberstein1967}).} \\
\hline
 $\alpha_n = 0.01 \frac{v+10}{\exp(\frac{v+10}{10})-1} $ & $\beta_n = 0.125 \exp(\frac{v}{80}) $ & $\frac{\mathrm{d} n}{\mathrm{d} t} = \alpha_n (1-n) - \beta_n n $ \\ 
\hline
 $\alpha_m = 0.1 \frac{v+25}{\exp(\frac{v+25}{10}) -1} $ & $\beta_m = 4 \exp(\frac{v}{18}) $ & $\frac{\mathrm{d} m}{\mathrm{d} t} = \alpha_m (1-m) - \beta_m m $ \\ 
\hline 
 $\alpha_h = 0.07 \exp(\frac{v}{20}) $ & $\beta_h = \frac{1}{\exp(\frac{v+30}{10})+1} $ & $\frac{\mathrm{d} h}{\mathrm{d} t} = \alpha_h (1-h) - \beta_h h $ \\
\hline
$h_0 = 0.596 $ & $ n_0 = 0.318 $ & $ m_0 = 0.052 $ \\
\hline
$C_a = 0.0 \left[\frac{\mu \mathrm{F}}{\mathrm{cm}^3}\right]$ & $C_m = 1 \left[\frac{\mu \mathrm{F}}{\mathrm{cm}^2}\right]$ & $R = 35.4 \left[\Omega \cdot \mathrm{cm}\right]$ \\
\hline
$\hat{g}_K = 36 \left[\frac{\mathrm{m.mho}}{\mathrm{cm}^2}\right]$ & $\hat{g}_{Na} = 120 \left[\frac{\mathrm{m.mho}}{\mathrm{cm}^2}\right]$ & $\hat{g}_l = 0.3 \left[\frac{\mathrm{m.mho}}{\mathrm{cm}^2}\right]$\\
\hline
$V_K = 12 \left[\mathrm{mV}\right]$ & $V_{Na} = -115 \left[\mathrm{mV}\right]$ & $V_l=-10.613 \left[\mathrm{mV}\right]$ \\
\hline
\end{tabular}
\end{center}
For the initial condition, we generate a narrow localised pulse for AP in the middle of the 1D space domain at time $t_0=0$ as
$v(x,t_0) = A_0 \mathrm{sech}^2 (B_0 \cdot x_0),$ where $x_0 = x - l_0 \cdot \pi,$
where $A_0$ is amplitude of the pulse, $B_0$ is the width parameter of the pulse, $l_0$ is number of 2$\pi$ sections in space. For axial current, we take zero initial value and values for $n,m,h$ are given in Table 1. As is usual with psuedospectral method, we are using periodic boundary conditions. 

For numerical solving of these models, the discrete Fourier transform (DFT) based (PSM) (see \cite{Fornberg1998,Raamat2021}) is used.
Variable $Z$ and its spatial derivatives can be represented with the help of DFT as
\begin{equation} \label{dft2}
\widehat{Z}(k,T) = \mathrm{F} \left[ Z \right]= \sum^{n-1}_{j=0}{Z(j \Delta X, T) \exp{\left(-\frac{2 \pi \mathrm{i} j k}{n} \right)}}, \quad \text{and,} \quad
\frac{\partial^{m} Z}{\partial X^{m}} = \mathrm{F}^{-1}\left[(\mathrm{i} k)^{m} \mathrm{F}(Z) \right],
\end{equation}
%
where $n$ is the number of space-grid points, $\Delta X=2 \pi/n$ is the space step, $k=0,\pm1,\pm2,\ldots,\pm(n/2-1),-n/2$; $\mathrm{i}$ is the imaginary unit, $\mathrm{F}$ denotes the DFT and $\mathrm{F}^{-1}$ denotes the inverse DFT. The idea of the PSM is to approximate space derivatives by making use of the DFT (see eqs.~\eqref{dft2})
reducing, therefore, the partial differential equation (PDE) to an ordinary differential equation (ODE) and then using standard ODE solvers for integration in time. However, in the Boussinesq \eqref{Mindlin} and iHJ \eqref{iHJ} equations, there are terms with mixed partial derivatives. To handle these, we rewrite equation \eqref{iHJ} (and \eqref{Mindlin}) so that all time and mixed partial derivatives are on the left-hand side of the equation and then do a change of variables to be able to use the PSM, see \cite{Raamat2021} Appendix A for details. An example Python script is given in \cite{Raamat2021} Appendix B, which can be used to solve \eqref{iHJ} by changing the initial condition and setting the coupling parameters and dissipation to zero. 

\printcredits

\bibliographystyle{model1-num-names}




\end{document}